\documentclass[fleqn,9pt]{SelfArx} 

\definecolor{color1}{RGB}{0,0,90} 
\definecolor{color2}{RGB}{0,20,20} 

\usepackage{hyperref} 
\usepackage{epstopdf}
\usepackage{siunitx}
\usepackage{cite}
\usepackage[figurename=Fig.,justification=justified]{caption}
\usepackage{pdfpages}
\AtEndDocument{\includepdf[pages={-},pagecommand={\thispagestyle{empty}},scale=1]{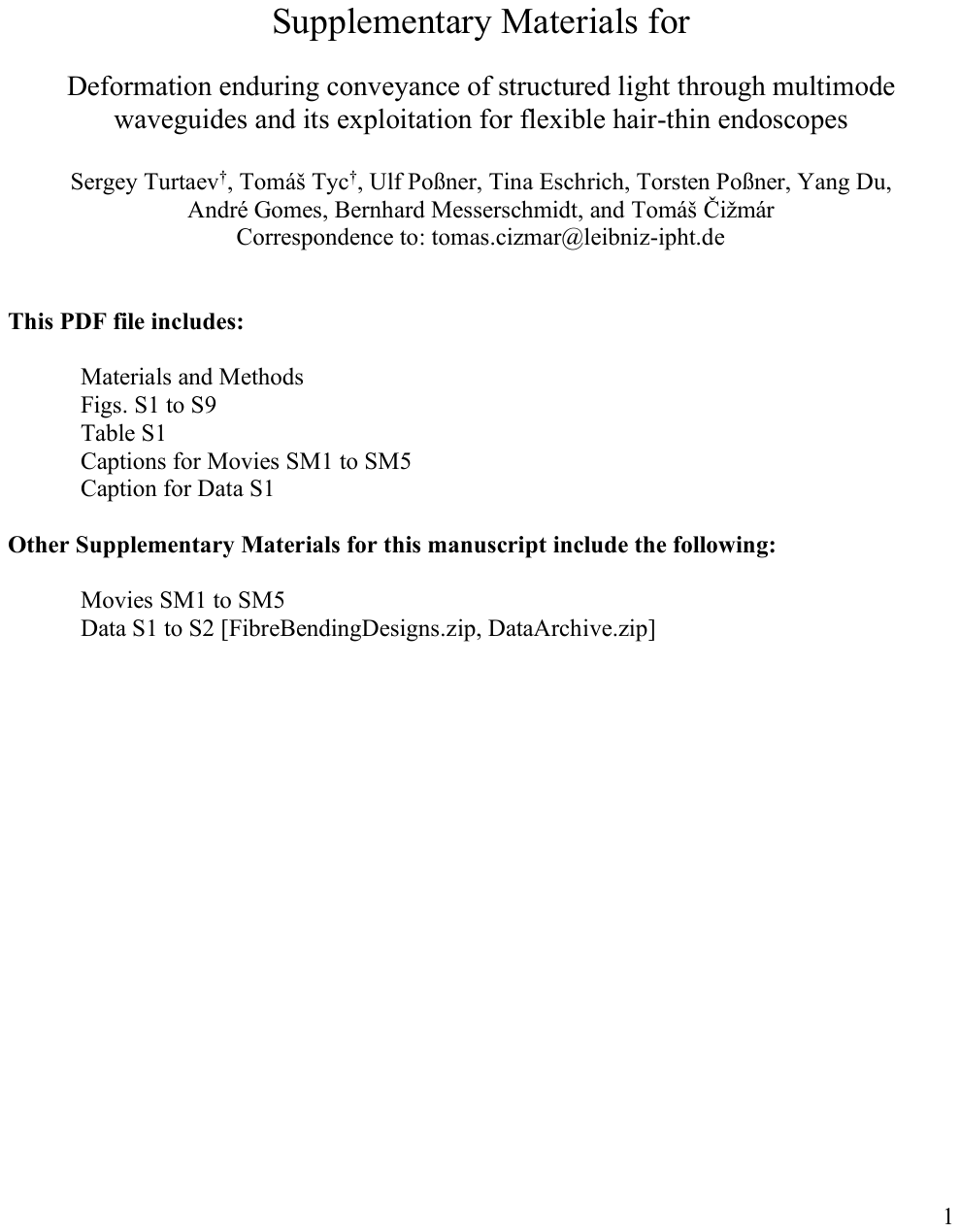}}

\hypersetup{hidelinks,colorlinks,breaklinks=true,urlcolor=color2,citecolor=color1,linkcolor=color1,bookmarksopen=false,pdftitle={Title},pdfauthor={Author}}

\JournalInfo{~} 
\Archive{} 

\PaperTitle{Deformation enduring conveyance of structured light through multimode waveguides and its exploitation for flexible hair-thin endoscopes} 

\Authors{Sergey Turtaev\textsuperscript{1,$\dagger$}, 
Tom\'{a}\v{s} Tyc\textsuperscript{2,3,$\dagger$}
Ulf Po{\ss}ner\textsuperscript{4}, Tina Eschrich\textsuperscript{1}, Torsten Po{\ss}ner\textsuperscript{4}, Yang Du\textsuperscript{1}, Andr\'{e} Gomes\textsuperscript{1}, Bernhard Messerschmidt\textsuperscript{4},  and Tom\'{a}\v{s} \v{C}i\v{z}m\'{a}r\textsuperscript{1,2,5,*}} 
\affiliation{\textsuperscript{1}\textit{Leibniz Institute of Photonic Technology, Albert-Einstein-Stra{\ss}e 9, 07745 Jena, Germany}} 
\affiliation{\textsuperscript{2}\textit{Institute of Scientific Instruments of CAS, Kr\'{a}lovopolsk\'{a} 147, 612 64, Brno, Czechia}} 
\affiliation{\textsuperscript{3}\textit{Department of Theoretical Physics and Astrophysics, Faculty of Science, 
Masaryk University, Kotl\'{a}\v{r}sk\'{a} 2, 61137 Brno, Czech Republic}} 
\affiliation{\textsuperscript{4}\textit{GRINTECH GmbH, Schillerstra{\ss}e 1, 07745 Jena Germany}}
\affiliation{\textsuperscript{5}\textit{Institute of Applied Optics, Friedrich Schiller University Jena, Fr\"{o}belstieg 1, 07743 Jena, Germany}} 
\affiliation{$^\dagger$\textbf{Authors contributed equally}}
\affiliation{*\textbf{Corresponding author}: tomas.cizmar@leibniz-ipht.de} 

\newcommand{\keywordname}{Keywords} 

\Abstract{\bf 
The remarkable advancements in our capacity to synthesise structured light have facilitated the generation of any desired optical landscapes and even controlling the spatial distribution of light propagating through optically complex media such as multimode fibres. The availability of precisely defined structured light at the extremity of an exceedingly narrow and flexible cable holds the potential to stimulate a diverse range of highly sought-after applications, encompassing rapid communication, quantum computing, and, notably, imaging. What we lack in reaching these aspirations is the resilience of such light transport to deformations of the waveguide. 
Although recent theoretical investigations have delineated the attributes of ideal multimode fibres capable of deformation-enduring conveyance of structured light, tangible fibres possessing this indispensable trait to a practical extent remain elusive.
Our study takes a deep dive into the precision of commercially available multimode fibres with the highest probability of demonstrating the phenomenon. 
We identified minuscule imperfections in their refractive index distribution, examined how these affect light transport when the fibre is deformed, and studied their implications for imaging applications. Our investigation has confirmed that these imperfections are indeed responsible for the undesirable alterations introduced into the output structured light fields during bending. Finally, as an alternative to standard graded-index fibres, manufactured by drawing silica-based preforms, we present narrow multimode waveguides in which the refractive-index profile has been established by ion exchange. These waveguides indeed exhibit previously unseen resilience of structured light transport even under severe deformation conditions and aptly fulfil the requirements of imaging applications.
}

\begin{document}

\fontsize{3.3mm}{4.3mm}\selectfont
\flushbottom 

\maketitle 


\thispagestyle{empty} 


\section*{Introduction} 

\addcontentsline{toc}{section}{Introduction} 
The recent explosion of research activity in the domain of structured light has already witnessed an impact across disciplines. The 2014 Nobel prize in Chemistry for the development of super-resolved fluorescence microscopy is probably the most famous example and numerous research and industrial applications nowadays emerge with steadily increasing pace. While we are able to routinely synthesise highly complex and vector structured light fields with millions of degrees of freedom, we lag behind in translating them from our bulky, table-top geometries into compact and integrated solutions and in their delivery to where they are needed. Exploiting multimode fibres (MMFs) to deliver structured light towards it application site or to convey it between remote sites of an optical system, could greatly accelerate these prospects.  
Methods to counteract the transformation of a structured light, which it experiences during propagation in a MMF, became very popular research domain in the past decade\cite{Bianchi:2012wc,Cizmar2012,Choi:2012tt,Papadopoulos:2013uk,Ploschner2015b}. MMFs are commonly considered as a special class of complex media. Transport of coherent light signals through them can therefore be described in conveniently chosen representation of light modes as a linear operator commonly referred to as the transmission matrix (TM)\cite{Popoff:2010to}. This relationship between input and output of the fibre must be known prior a system utilising MMF can be used for synthesis of desired structured light outputs. In most cases it is obtained empirically in procedure referred to as the calibration \cite{Cizmar2011}. As the calibration takes place, one needs full optical access to both ends of the MMF segment and it typically involves sequential propagation of a complete basis of input modes through the MMF while monitoring the output fields with the use of interferometry. When the TM becomes available, one can synthesise any desired  optical field at the output (distal) end by generating the corresponding input field and coupling it into the input (proximal) end of the MMF. The optimal results are reached when one controls the spatial distribution of all aspects of the input field including its amplitude, phase and polarisation\cite{Gomes:22}.
Even in the highest performing systems reported so far, the calibration can be completed in a few minutes.
Arguably, the highest application potential of this approach has been found in imaging. The unparalleled information density of MMFs allows them to convey images through hair-thin instruments having orders of magnitude smaller footprint when compared to other established technologies. Advanced geometries have already enabled deep-tissue observations of fluorescently labelled neurones in living animal models\cite{Ohayon2017,Turtaev2018,Vasquez-Lopez2018,Stiburek:2023nr}, label-free non-linear imaging of biological tissues\cite{Cifuentes2021}, minimal access observations of macroscopic objects\cite{Leite2021} which was also combined with time-of-flight based 3D reconstruction\cite{doi:10.1126/science.abl3771}.  Further applications beyond imaging include delivery of structured light forming dynamic 3D holographic optical tweezers\cite{Leite2018} and micro-manufacturing\cite{morales2017three,delrot2018single,konstantinou2023improved}.

With the calibration procedure completed, the application relying on delivery of a structured light can proceed in principle indefinitely, as long as the layout of the fibre remains unchanged. Initial experiments have shown that the geometries might tolerate a small amount of MMF bending, stronger MMF deformations have, however, resulted in significant changes to the TM and a fresh calibration was required to retain the original performance.   
This constraint to stationary MMF geometries has been seen as the most significant drawback of this perspective method from its inception. The exemplar case of \emph{in-vivo} neuroscience would greatly benefit from the possibility of deep-tissue minimum access observations in brains of fully functioning motile animal models, yet because of this very problem, all uses of MMF-based imaging demonstrations have been restricted to immobilised animals under anaesthesia where important processes behind neuronal signalling are significantly suppressed or completely silenced. This, combined with analogous desires within numerous other application areas  has triggered the attention of many, who within the recent years have attempted to provide a solution. One group of possible strategies explore the growing power of computing and electronics. These involve repeated calibration for all intended fibre layouts upfront\cite{Farahi:13,Wen:2023vc}, predicting changes to TM based on the observed fibre layout\cite{Ploschner2015}, eliminating the need for optical access to the MMF's distal end \cite{PhysRevX.9.041050,Wen:2023vc,Li2021,Li2021a} and the exploitation of deep learning \cite{Fan:19,Zhao_2021,Abdulaziz:2023fs}. 

Separate efforts concentrated on the MMF itself. It has been identified very early that TMs of different kinds of MMFs from various manufactures exhibit very different level of resilience to bending\cite{Loterie2017}. Broader theoretical insight into light transport through MMFs have predicted that graded index fibres following perfectly parabolic distribution of the square of the refractive index ($n^2$) across its core shall feature the most deformation-enduring conveyance of structured light (DECStL)\cite{BoonzajerFlaes2018a}. 
The phenomenon of DECStL can be best outlined when considering photons as particles. The specific distribution of refractive index profile forms a harmonic potential well\cite{leonhardt2012geometry} while a bend of the fibre provides linear potential slant. 
A photon moving inside an adiabatically curved fibre thereby still experiences almost identical harmonic potential well, albeit slightly shifted towards the outer side of the bend.  Its evolution therefore does not deviate from that in a straight fibre.
%
With the steep progress in quality of commercially available graded index fibres, namely the recent introduction of the OM5 standards (where the extended bandwidth indicates ever-highest precision in the refractive index distribution), several crucially important questions arise: To what extent is DECStL available in fibres commercially available nowadays? What precision would be required to meet the application demands without compromises? Can better-performing fibres be produced by any currently available technology? In this paper we consolidate all necessary requisites to answer these questions. 

We quantify deviations from the ideal parabolic refractive index distributions across a spectrum of the most promising commercially available candidates within our reach. We develop rigorous theoretical model predicting light transport through such aberrated fibres and simulate their imaging performance under bending deformations. Further, by benchmarking these predictions against the direct experimental reality, we confirm that minuscule manufacturing aberrations, typically in the order of $10^{-5}$ -- $10^{-4}$, are indeed to blame for the changes to the TM induced by fibre bending.    

Most importantly we propose and implement alternative manufacturing approach based on ion exchange\cite{Messerschmidt:95}, which is expected to improve the precision of refractive index distribution and reach characteristics more favourable for bending resilience when compared to standard graded-index fibre manufacturing methods. The novel waveguides have been designed to allow propagation of $\approx$17 000 modes, which could be used to convey images with the ever-highest pixel resolution\cite{Li2021a}.  
The resulting products, referred to here as the DECStL waveguides exhibit minimal imaging performance degradation even when deformed all the way towards their damage limits, which contrasts widely with the results achieved while using commercially available fibres. 

\section*{Results}
\addcontentsline{toc}{section}{Results} 

\begin{figure*}[h!] 
   \includegraphics[width=\textwidth]{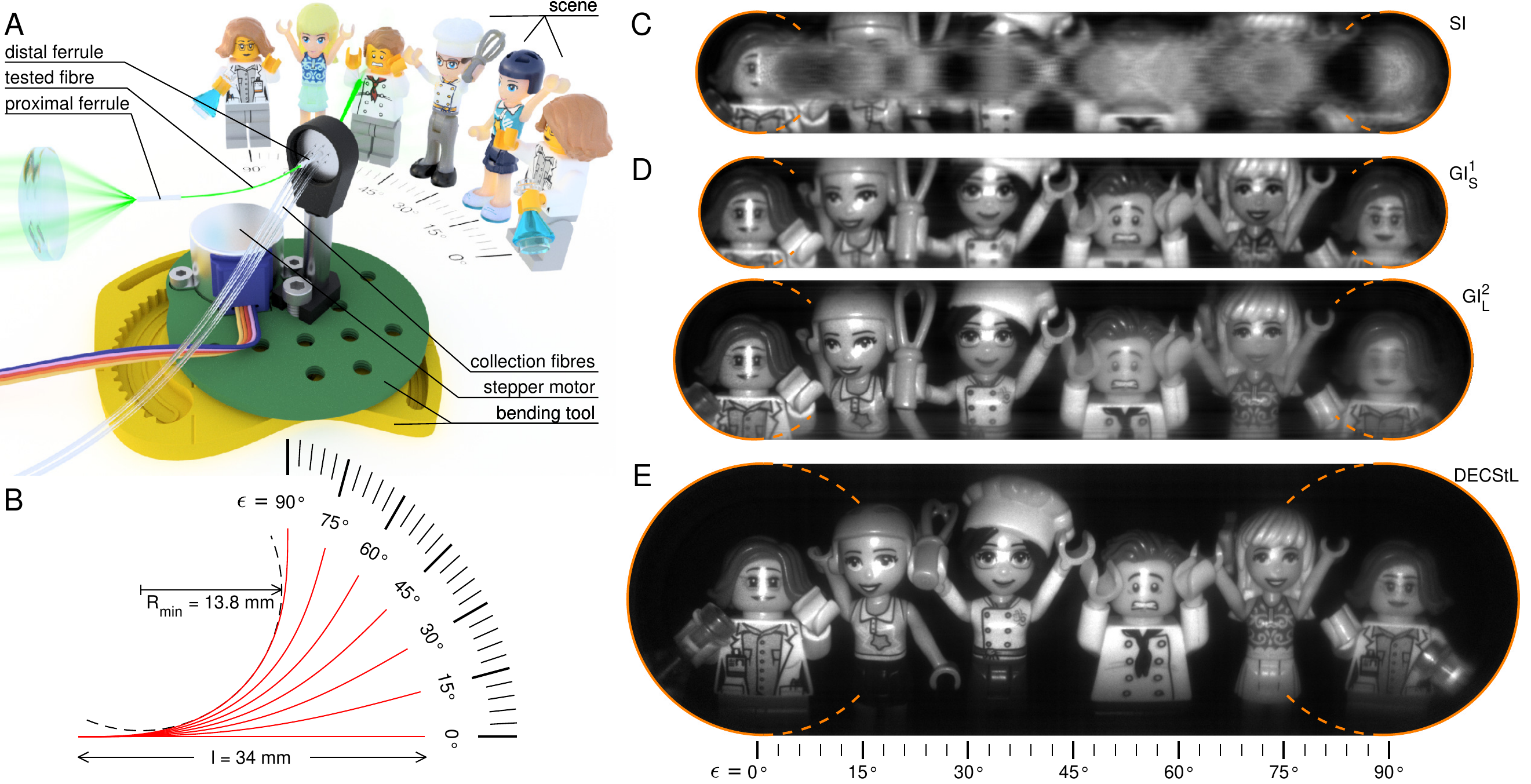} 
   \caption{{\bf Qualitative illustration of bending resilience.} ({\bf A}) Bending tool translating and reorienting the distal end imposing adiabatic curvature along the used fibre. ({\bf B}) Range of adiabatic fibre layouts used. ({\bf C}) Scene recorded by SI fibre. The central part is obtained by combining narrow middle sectors (5 pixels wide) of images obtained as the fibre was bend across the full range of the bending tool. Half-moon parts at the left and the right are taken from the first and the last recorded image respectively. ({\bf D}) Scene recorded by the best performing commercially available graded-index fibres of our selection and the best roll orientation, with the lowest and the highest information capacity ($\mathrm{GI_{S}^{1}}$ and $\mathrm{GI_{L}^{2}}$). ({\bf E}) Scene recorded by DECStL waveguide for its full information capacity. Roll orientation played no observable role.}
   \label{fig1}
\end{figure*}
For the experimental studies, six most promising commercially available candidates for bending-resilient fibre, featuring different core sizes and numerical apertures, have been selected based on our market research. 
Their identifiers in this work ($\mathrm{GI_{S/M/L}^{1/2}}$),
manufacturers and specifications are explained in supplementary materials (SM, table S1). This selection has been complemented by a step-index fibre (SI) which has been most commonly used in previous experimental work on this topic. 

The last candidate to be assessed for its bending resilience is the newly developed DECStL waveguide. A cane from alumino-boro-silicate glass\cite{glasspatent} has been drawn in an in-house fibre-drawing tower into the diameter \SI{125}{\micro\metre} (matching that of all the fibres used) and subjected to ion exchange as detailed in SM, aiming to produce parabolic distribution of $n^2$ giving the NA of 0.5. The available ion-exchange reactor has enabled lengths of up to \SI{54}{\milli\metre}, therefore all candidates have been tested at matching length. 

For all experiments we use the geometry of the far-field holographic endoscope \cite{Leite2021,doi:10.1126/science.abl3771}, detailed in SM. It has been optimised to adapt to optical fibres with core size ranging from \SI{50}{\micro\metre}  to \SI{125}{\micro\metre} and NAs between 0.2 and 0.5. 

In all our studies the fibre's TM was measured at the initial, almost straight layout. A small contortion (negligible with respect to other bent layouts under investigation) has been allowed to avoid stretching the fibre at this layout.
The instrument is equipped with custom-made bending tool, which has been designed to introduce strictly adiabatic bending deformation to the \SI{34}{\milli\metre} long segment of the used fibre (remaining length when using two \SI{10}{\milli\metre} long termination ferrules) by simultaneous positioning and orienting of the distal end. While there are numerous other possible parametrisable layouts of the fibre, which can be introduced by positioning and orienting the segment's ends, the adiabatic shape featuring smooth curvature and its first derivative is the only one that requires no torques applied to the ends (see SM for details) and therefore it is the easiest shape to reproduce without imposing other means of deformations such as local stress. Following the TM measurement, the bending tool has been employed. It  introduces any bend ranging from the initial (calibrated) layout and the terminal layout with the distal end yaw ($\epsilon$, angle of the distal end axis measured from the proximal end axis) reaching \SI{90}{\degree} and extreme radius of curvature reaching \SI{13.8} {\milli\metre} (see Fig. \ref{fig1}B).    
Imaging is achieved through point-by-point scanning of a diffraction-limited spot over the objects within the field of view, while collecting the back-scattered light intensity by additional fibres. As discussed elsewhere\cite{Leite2021}, it is very challenging to implement both, object illumination and collection of returning signals through the same fibre in the far-field reflectance geometry, mainly due to overwhelmingly strong reflections at the facets and scattering at defects of the fibre. The issue vanishes when imaging in close proximity of the distal end as well as while employing fluorescence-based\cite{Stiburek:2023nr}  or non-linear\cite{tragaardh2019label} imaging approaches where respective excitation and signal collection optical pathways can be separated spectrally. While previous works used the narrowest possible collection fibres in the closest possible proximity of the illumination waveguide in order to maintain minimal dimensions of the instrument, our experiment settings are optimised purely for studies of the bending resilience. Therefore we include six step-index collection fibres of much larger cores (\SI{400}{\micro\metre}) and NAs (0.5), each displaced several millimetres from the illumination fibre under investigation. Next to considerably stronger signals, this solution does not affect the fibre layout and avoids image contamination by speckle artefact originating from the coherent nature of the signal\cite{Leite2021}.     

\begin{figure*}[h!] 
   \includegraphics[width=\textwidth]{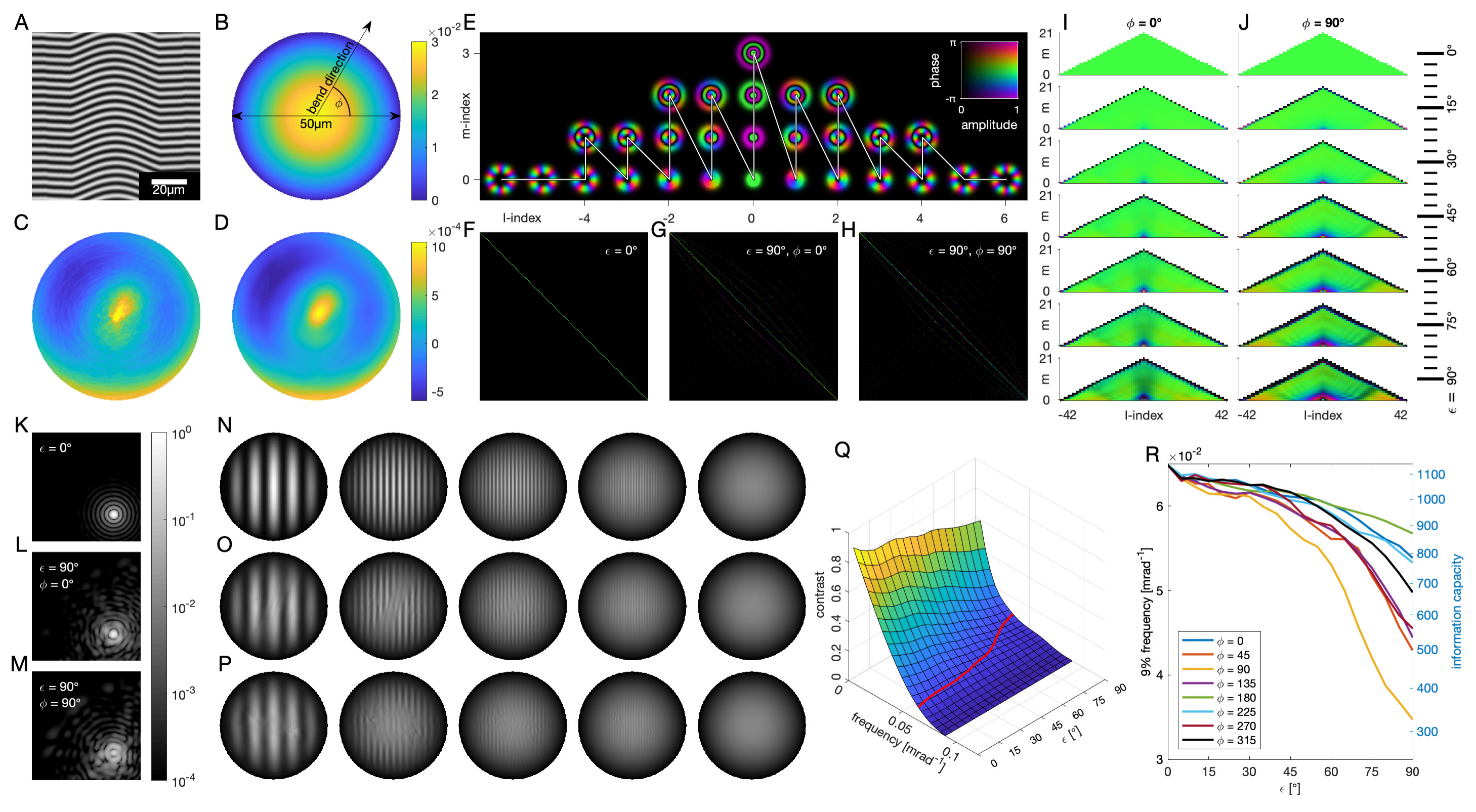} 
   \caption{{\bf Imaging performance bending resilience modelling for aberrated fibres.} $\mathrm{GI_{M}^{2}}$ is used as model case for all data shown in this figure. ({\bf A}) Interferogram of fibre placed perpendicularly with respect to the object wave direction. ({\bf B}) Computer tomography-based reconstruction of refractive index difference from that of the cladding material. ({\bf C}) Isolated aberration after paraboloid dependence has been subtracted. ({\bf D}) Aberration after low-pass filtering by projection onto 300 lowest Zernike polynomials. ({\bf E}) PIMs of perfect parabolic fibre. Solid line indicates their order considered in deformation operators below. ({\bf F} to {\bf H}) DOs for straight fibre (unitary matrix) and two cases bent fibre (distal yaw of \SI{90}{\degree}) for different roll orientations. For better clarity the basis in these examples has been reduced to 320 PIMs of the lowest order (largest propagation constants), where the strongest effects of bending appear. ({\bf I} and {\bf J}) Diagonal components of complete DOs organised into mode pyramids (as in {\bf D}) for two different roll orientations and gradually increasing distal end yaw. ({\bf K} to {\bf M}) Examples of far-field foci formed through deformations of {\bf F} to {\bf H}, considering the fibre was calibrated at the straight layout. ({\bf N} to {\bf P}) Simulated raster-scanning-based imaging of single-spatial-frequency gratings for corresponding deformations. ({\bf Q}) Contrast transfer function evolution under the influence of bending. Values of contrast were 
averaged over the field of view and four different orientations of the grating's fringes. The red contour signifies the resolution limiting 9\% frequency (analogous to Rayleigh criterion). ({\bf R}) 9\% frequency and total information capacity dependence on the distal end yaw for several different roll orientations of the fibre. }
   \label{fig2}
\end{figure*}

Prior to our detailed quantitative assessment of the bending resilience we provide a qualitative demonstration of all used fibre types by capturing the same scene while increasing the distal end yaw (see Fig. \ref{fig1}A). In most commercially available fibres the refractive index aberrations are not azimuthally independent, which results in significantly diverse bending resilience for different roll orientations of the fibre ($\phi$). In Supplementary movie SM1, we provide records of the scene during progressive bending for the worst as well as the best performing roll orientation. The DECStL waveguide exhibited no significant differences for different roll orientations, therefore only one orientation is provided. The study however includes imaging results for reduced information capacities (virtually reducing the NA and the core size of the in-coupled signal) to allow performance comparison with the commercial fibres. 
A selection of the results is visualised in Fig.~\ref{fig1}, for the step-index fibre (Fig.~\ref{fig1}C), two examples of the commercial fibres performing at the lowest and the highest information capacities available, both at their best-performing roll orientation (Fig.~\ref{fig1}D), and the DECStL waveguide at its full capacity (Fig.~\ref{fig1}E).

Our quantitative investigation is conceived to provide the rigorous explanation for the light transport changes and the image degradation under the influence of bending. 
The hypothesis, which we put forward for the observed lack of robust bending resilience in common graded-index fibres is the presence of perturbations in their refractive index profile. In the following, we therefore measure the refractive index profile, predict the bending resilience by a numerical model and compare the result with the experimental reality. 

The refractive index profile (its difference from the cladding material) is acquired by the means of computer tomography, using records from multidirectional interferometric measurements (see Fig.~\ref{fig2}A for an example of an interferogram acquired and SM for further details). Subtracting the ideal fibre profile of the optimum fit for the centre of symmetry and NA reveals the identified aberrations (see Fig.~\ref{fig2}C), exhibiting visible high-frequency wavelets. Using the raw data in our numerical model resulted in much stronger inter-mode coupling than that observed in our experiments, therefore we assumed that the high frequency wavelets represent numerical artefacts of the computer tomography procedure. We therefore low-pass filtered the measurements by decomposition into the Zernike polynomials defined at the area of the fibre core (value given by manufacturer's specifications) and use only these of the first 24 orders (300 elements, see Fig.~\ref{fig2}D), which represents the largest possible threshold before the high-frequency artefacts start to emerge. The measurement results from all the used fibres are presented in SM.

Our numerical model to predict the light transport through perturbed fibres assumes that the refractive index distribution is close to the ideal shape, featuring an arbitrary but relatively small perturbation, which does not change along the fibre's axis. The model is built upon the identification of modes in uniformly bent and perturbed fibres, with the transmission through arbitrarily curved fibres being realised in discrete segments, in each of which the curvature can be regarded as essentially uniform. The modes of the perturbed fibre are expressed as a superposition of the propagation invariant Laguerre-Gaussian modes (PIMs, see Fig.~\ref{fig2}E) of the unperturbed fibre. The model is detailed in supplementary information, together with discussions about its most important predictions.   These include the realisation that perturbations which are symmetric with respect to the fibre axis (point symmetry of refractive index cross-section) cause much weaker changes to light transport when compared to the antisymmetric ones. 
Further, the model predicts that the influence of any given perturbation grows proportionally to fibre diameter and it falls with the square of the NA.   
The result of the model is a deformation operator (DO), which is defined as a matrix, with which one must multiply the TM of the straight fibre in order to obtain that of the bent fibre. Figs.~\ref{fig2}G~and~\ref{fig2}H exemplifies DOs in their matrix form, while Figs.~\ref{fig2}I~and~\ref{fig2}J show evolution of the DO's diagonal components (PIMs) under the influence of increasing amount of bending for two different roll orientations of the fibre.  
To predict the imaging performance, we simulate imaging by virtually exposing grating targets of varying spatial frequencies (sinusoidal) by sequence of farfield foci, one for each desired pixel.  
For each we identify the superposition of PIMs which is to be coupled into the straight fibre in order to achieve the desired diffraction-limited focus (see Fig.~\ref{fig2}K).  Analogous illumination fields of the bent fibre is obtained by multiplying the same vector of PIMs with DO calculated by our theoretical model for the actual fibre layout (see Figs.~\ref{fig2}L~and~\ref{fig2}M). Each pixel of the resulting image is obtained as the inner product of the intensity distribution of the corresponding focus with the grating target (see figs.~\ref{fig2}~n-p). An example of contrast of the resulting images, averaged over the field of view and four different orientations of the grating targets is presented in Fig.~\ref{fig2}Q as a function of the target's spatial frequency and the amount of bending. Following the Rayleigh definition of resolution, we consider the limiting spatial frequency at the contrast level of 9\%\cite{Leite2021}. The total number of resolvable features (information capacity) can be estimated as square of the limiting frequency multiplied by the area of the field of view. These quantities are presented in Fig.~\ref{fig2}R for 8 different roll orientations of the fibre.

\begin{figure*}[htbp] 
   \includegraphics[width=\textwidth]{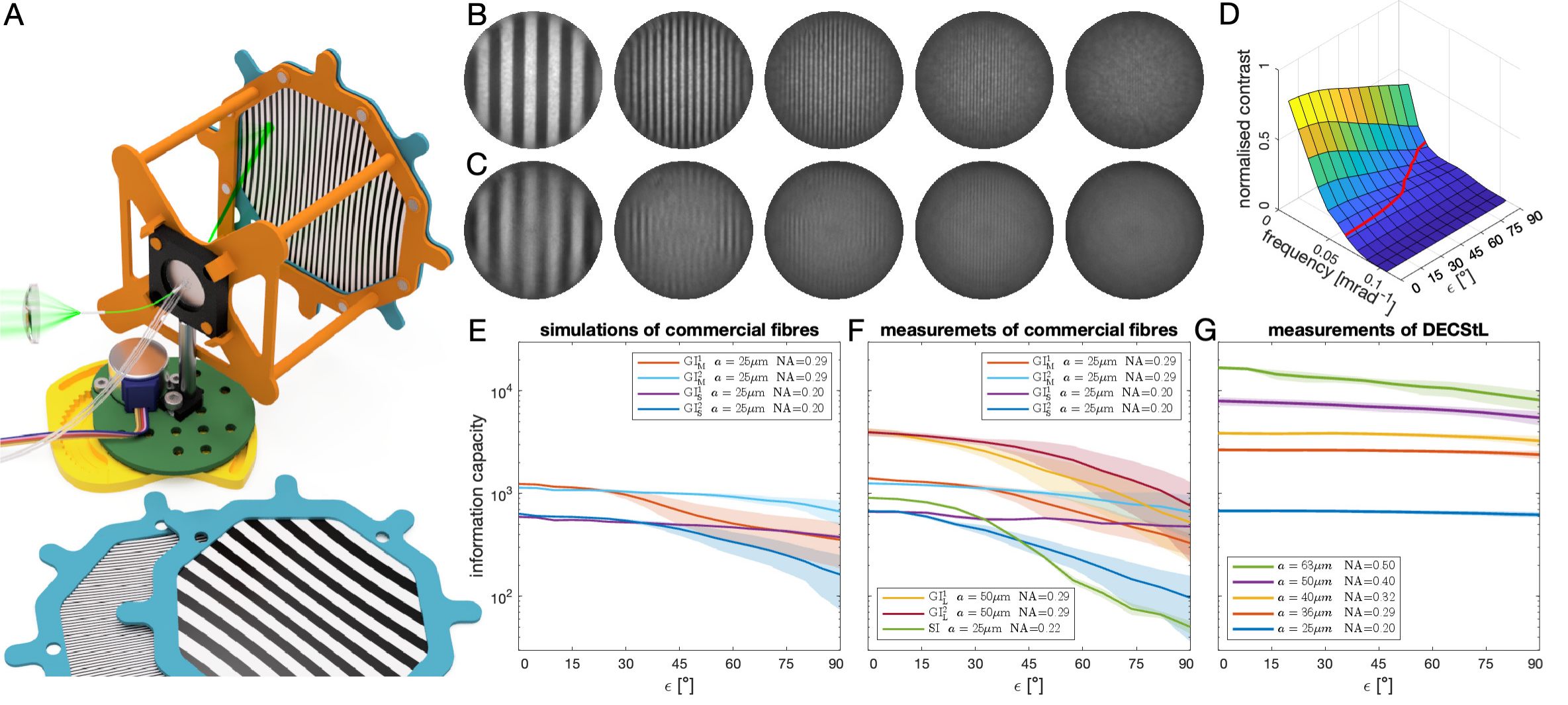} 
   \caption{{\bf Experimental assessment of imaging performance resilience to bending and its comparison to theoretical prediction.} $\mathrm{GI_{M}^{2}}$ is used as model case for results shown in panels {\bf A} to {\bf C}. ({\bf A}) Illustration of experimental settings and procedure. ({\bf B} and {\bf C}) Raw images of Ronchi targets obtained for straight and bent (\SI{90}{\degree} distal yaw) fibre layouts respectively.  ({\bf D}) Normalised contrast transfer function evolution under the influence of bending. The red contour signifies the resolution limiting 9\% frequency. ({\bf E})  Modelled decline of total information capacity (number of resolvable features) under the influence of bending for four fibres which our model could handle. Shaded intervals signify spread obtained for different roll orientations. ({\bf F}) Corresponding experimental results for the same fibre selection as in {\bf E}, with additional fibres of higher capacity and the step-index type.  ({\bf G}) Equivalent measurements on DECStL waveguide with information capacity of the initial layout manipulated by aperturing the input signals.}
   \label{fig3}
\end{figure*}

Finally we provide direct experimental evidence quantifying the fibres' bending resilience. The bending tool was equipped with holder carrying frequency targets, which are kept stationary with the distal end of the used MMF (see Fig.~\ref{fig3}A). The square (Ronchi) targets of various frequencies are printed by black ink on a white paper and glued onto magnetic frames allowing four different orientations of the fringes with respect to the system. Following the calibration at the initial close-to-straight layout of the fibre the system provides images along the full interval of adiabatic contortions for variety of spatial frequencies and four orientations of the fringes (see Fig.~\ref{fig3}B and \ref{fig3}C for examples). The measurements were repeated for six roll orientations of the used fibre and all fibres considered. Analogously to the previous simulation approach, the images are used to quantify the contrast. There are three experimental issues, influencing the measured values of contrast. First, due to presence of noise in TM measurement and the used light-modulation mechanism, our diffraction-limited foci contain only around 70\% of the total transmitted power, the fraction varies for different fibres with the dimension of their TMs. The remaining power is distributed in the form of speckled background across the field of view. Further, the black ink is not perfectly absorbing. Lastly, the Ronchi targets used had square rather than sinusoidal reflectivity cross-section. In all the cases, however the contrast, obtained from the Fourier transform of the image, was affected for all frequencies of the same dataset by the same factor. The results are therefore normalised, demanding to reach unity when extrapolating the obtained dependency for of the initial layout to the spatial frequency of zero (see Fig.~\ref{fig3}C) as expected in the case free from the above experimental issues.         

The side-by-side comparison of the obtained results is summarised in Figs.~\ref{fig3}E to \ref{fig3}G in the form of information capacity reduction under the influence of bending. The shaded areas represent confidence intervals ($\pm$ one standard deviation) for results obtained at different roll orientations while the solid lines show the average.  The simulations, which closely followed the experimental conditions (see Fig.~\ref{fig3}E), could only be completed for four fibres of the low and medium information capacity as the available computing power (96 CPU cores, 2TB memory) was insufficient to handle larger datasets efficiently. The step-index fibre was also not simulated as the difference of the step-index refractive index profile from the ideal paraboloid is too high to be considered as a perturbation. Yet, direct comparison with the experimental results shown in Fig.~\ref{fig3}F reveals very similar trends for the corresponding fibres in both, the mean values and the variations across different roll orientations. Clearly, $\mathrm{GI_S^1}$ (Draka WideCap-OM5) exhibits the highest bending resilience, which corresponds well with the qualitative impressions introduced earlier.  Unfortunately, it offers only a very small information capacity. Higher-capacity commercial fibres exhibit much steeper performance degradation under equivalent bending. Analogous measurements of the DECStL waveguide shown in Fig.~\ref{fig3}G however reveal immensely enhanced bending resilience, greatly outperforming the best tested commercial fibre even when used to convey orders of magnitude larger amount of imaging information.

\section*{Discussion}

In summary, our studies show that, with sufficiently precise profile of refractive index distribution, deformation-enduring conveyance of structured light through multimode fibre optics can be reached to unprecedented levels. 

The work provides detailed analysis of refractive index perturbations seen in commercially available graded-index fibres and it links them with the undesired changes to light transport taking place while fibre bending. 
Our theoretical model, verified by direct comparison with the experimental reality allows one to predict bending-resilience of a given fibre, based purely on the knowledge of its refractive index profile. 
The model predicts that symmetric and antisymmetric perturbations have starkly contrasting influence of light transport, which brings about interesting implications for the fibre manufacturing processes. 
To appreciate its importance, consider the example of $GI_s^1$ and $GI_s^2$, which have a very similar parameters and also the refractive index perturbations. Yet $GI_s^1$ features slightly asymmetric profile with respect to the axis, while $GI_s^2$ is almost perfectly symmetric (see supplementary figure S3).  Yet, both the numerical model-based prediction and the experimental reality concordantly exhibit very differing bending resilience outcomes for these fibres. 
Investing effort to eliminating all radially symmetric perturbations but also the ellipticity of the fibre is therefore not advisable when striving to improve the bending resilience, one should rather focus mainly on making the production process as symmetric with respect to the fibre axis as possible.
Further, the model predicts that same perturbation may affect the bending resilience very differently shall it be present on fibres with different core size and NA. The most simple way to support bending resilience in practical cases is to opt for the highest NA fibre with the smallest possible size of its core. 
These considerations allows us to define a simple quantity which is available purely from the RI profile measurement. The bending resilience predictor (BRP), introduced in SM  (Materials and Methods, part 7) it is insensitive to fibre parameters and it can serve as a rough indication wether or not the waveguide in question is suitable as bending resilient medium. The relative information capacity of the bent waveguides (related to the performance of the original straight fibre for which the TM has been acquired), as a function of BRP, is shown in SM (supplementary figure S7) for all our experimental results and simulations. We believe that more robust version of this study could soon form the basis for a wide-spread quality standard to describe MMF bending resilience. 

Despite a very satisfying prognosis on bending resilience, a notable disparity exists between the model's projections and the empirical findings. The model anticipates the preservation of circular polarization state in light propagation, both within straight and bent fibre configurations. This has been empirically validated in step-index fibers\cite{Ploschner2015}, however, graded-index fibres exhibit a distinct response, resulting in the complete randomisation of polarisation states at the output. The underlying mechanisms for this dissimilarity remain unclear, and it will be a subject of our future research agenda.

We present novel approach for manufacturing flexible multimode waveguides based on ion exchange, providing very high degree of bending resilience and simultaneously featuring very high information capacities, combination of which significantly exceeds that of commercially available competitors. As seen in the measurements of refractive index profiles, its absolute perturbation is comparable to that found in the best-performing commercial fibres. Yet, as we explain in SM (Materials and Methods, part 3), their magnitude must be weighted against the square of the NA in order to predict their bending resilience. 
The success of the DECStL waveguide therefore results from both, the low perturbations and the significant NA enhancement, both enabled by the ion-exchange process.

Our work is likely to rise numerous questions which cannot be answered within the scope of this paper at the depth of the above presented results. Yet there are several important findings which are worth outlining:

As we performed our studies with relatively short fibre segments, it is desirable to know how the bending resilience translates to longer ones. In order to provide an initial insight, we provide comparable study at \SI{150}{\milli\metre} long fibre segment in SM (Materials and Methods, part 11) and showcase imaging of a scene with \SI{1}{\metre} long fibres in Supplementary Movies SM2 and SM3. These results clearly show that bending resilience is significantly enhanced for longer fibres whose distal end reaches a certain yaw angle along paths or smaller curvature. 

Our study has not covered twisting deformations in detail. We argue that most of conceivable applications could be engineered minimising the need for twist deformations, therefore it is of much lower importance when compared to bending.  
When applying a similar derivation approach to the one used for analysing the effects of bending on perturbed fibres, one might find that dependence on core size and NA holds also for twist deformations, but the significance of symmetric and antisymmetric perturbations reverses. Our supplementary movie SM4 shows that small amount of twist affects the DECStL waveguide's performance minimally, while it can play a profound role in commercial fibres.

We anticipate that a novel class of multimode waveguides, capable of achieving or surpassing the performance demonstrated by the DECStL waveguide described herein, will have broad applications across diverse research and industrial sectors. Spatially structured light fields play a crucial role in numerous manufacturing processes, demanding meticulous attention to maintaining the alignment of individual components. By constructing optical pathways with components that can be macroscopically translated without impacting the resulting structured fields, substantial reductions in system complexity and costs may be realized.
Telecommunications, in particular, stands to benefit significantly from the utilisation of bending-resilient fibers, enhancing spatial domain multiplexed data transfer. Additionally, within the realm of quantum computing, DECStL MMFs will facilitate the delivery of quantum optical states between remote processing units.
Primarily motivated by imaging applications, our work foresees that the adoption of DECStL waveguides will address the most significant limitation of MMF-based holographic endoscopes. This breakthrough paves the way for their rapid integration into various promising fields, including \emph{in-vivo} neuroscience. In this context, these fibers offer the unprecedented capability to enable high-resolution structural imaging of deep-seated brain structures in motile and behaving animal models.
We anticipate that our work will stimulate fibre producers to include the BRP or other unified DECStL quantifying standard of their products amongst their specifications and focus on development of new, further empowered DECStL waveguides.  

\bibliographystyle{nature}

\end{document}